\documentclass[sigconf]{acmart}
\usepackage[table,xcdraw]{xcolor}
\AtBeginDocument{%
  \providecommand\BibTeX{{%
    \normalfont B\kern-0.5em{\scshape i\kern-0.25em b}\kern-0.8em\TeX}}}

\setcopyright{acmcopyright}
\copyrightyear{2018}
\acmYear{2018}
\acmDOI{10.1145/1122445.1122456}

\acmConference[Woodstock '18]{Woodstock '18: ACM Symposium on Neural
  Gaze Detection}{June 03--05, 2018}{Woodstock, NY}
\acmBooktitle{Woodstock '18: ACM Symposium on Neural Gaze Detection,
  June 03--05, 2018, Woodstock, NY}
\acmPrice{15.00}
\acmISBN{978-1-4503-XXXX-X/18/06}

\begin{document}

\title{Trust as a Metric for Resiliency in Signed Social Networks}

\author{Harsh Patel}
\authornote{Both authors contributed equally to this research.}
\email{harsh.patel@iitgn.ac.in}
\affiliation{%
  \institution{Indian Institute of Technology Gandhinagar}
  \country{India}
  }

\author{Shivam Sahni}
\authornotemark[1]
\email{shivam.sahni@iitgn.ac.in}
\affiliation{%
  \institution{Indian Institute of Technology Gandhinagar}
    \country{India}
}
\author{Pushkar Mujumdar}
\email{pushkar.mujumdar@iitgn.ac.in}
\affiliation{%
  \institution{Indian Institute of Technology Gandhinagar}
  \country{India}
}

\renewcommand{\shortauthors}{Patel and Sahni, et al.}

\begin{abstract}

Recent technological advancements have resulted in a surge in online trading, raising severe concerns about theft and fraud, especially on platforms like {\em Bitcoin OTC} (over-the-counter), where users' identities remain anonymous. To mitigate the risk, it has become essential to capture the reputation of users based on their trade histories. The who-trusts-whom signed network of people has the capability to reflect the nature of such positive and negative relations between the users. It can be used to analyze linkage patterns, strength, and resiliency of such platforms. Due to the dynamic nature of trust between individuals, these trust networks are often vulnerable to link or node failures, making it critical to understand the stability of such systems.
In this paper, we consider the problem of quantifying the resiliency of signed networks with the help of trustworthy community structures.
We propose a metric for computing the {\em Trustworthiness} of a community structure. Using the trustworthiness scores of all communities structures, we generate a pipeline for assessing the resiliency of a signed network. We also show how these generated resiliency scores are concordant with the true nature of the network.
\end{abstract}

\ccsdesc[500]{Information systems~Clustering}
\ccsdesc[500]{Information systems~Reputation systems}
\ccsdesc[100]{Information systems~Data analytics}
\ccsdesc[100]{Information systems~Social tagging systems}
\keywords{signed networks, resiliency, community detection, link prediction}


\maketitle

\section{Introduction}

Resilience is the ability of a complex system to withstand the disturbances caused due to any internal failure or natural changes and recover quickly to maintain its functionality. Understanding how complex networks demonstrate resilience has become quite critical because of the increasing chances of crises in various systems like transportation, communication, and finance \cite{Giannoccaro2018}.

Signed networks are a powerful way of data representation that effortlessly reflect the mixture of positive and negative relations among entities. These signed networks are even used to capture weighted social relationships between people, like the strength of trust/mistrust between two people on a particular scale. In this paper, we tackle the problem of quantifying the trustworthiness of individual communities and the resiliency of the overall network by utilizing the signed trust relationships in the network.  

The interplay between the positive and negative edges of a signed network leads to the formation of conflicting communities. Thus, it has become essential to exploit the signed information to identify more stable communities to understand the strength of these networks. There have been previous works like \cite{9306909}, and \cite{10.1145/1753326.1753532}, focusing on measuring the stability of a community in signed social networks.

Transaction networks like the Bitcoin OTC are often incomplete graphs as there may be tacit relations of trust/mistrust that do not appear in the graph. The problem of predicting weights of such links is quite significant and has been well explored by the community using approaches like random-walk based \cite{Wan2019}, and generative adversarial network (GAN) framework based \cite{10.1145/3372923.3404805}. In our work, we use the Fairness-Goodness approach developed in \cite{7837846}, which uses the {\em Social Balance Theory} \cite{doi:10.1080/00223980.1946.9917275} to predict the unknown links in the network. According to the Social Balance Theory, ``A friend of an enemy is also an enemy''. The work by \cite{Facchetti20953} shows that currently available large online social networks are concordant with the social balance theory. This edge weight prediction is often helpful in improving the traditional tasks in real-world signed networks like network analysis \cite{10.1145/1753326.1753532} and community detection \cite{PhysRevE.80.036115}. 

We use the predicted edge weights in order to determine a {\em trust-worthiness} score of a community. Community detection plays a significant role in understanding the underlying patterns in complex networks. However, it has been significantly under-explored for signed networks as compared to unsigned ones \cite{Esmailian2015}. Few previous works include modified Spectral clustering \cite{doi:10.1137/1.9781611972801.49}, Correlation clustering \cite{Bansal2004} for community detection in signed networks. 

Using such graph clustering approaches over the Bitcoin OTC Trust Network, we identify the trustworthiness of the communities in this transaction network. This trustworthiness metric is further used to quantify resiliency. We show that these resiliency scores are a good estimate for understanding the stability of the network. With this work, we present a significant base for further study of resiliency quantification problem in signed social networks. 

\section{Dataset}

We use the Bitcoin OTC Trust Weighted Network \cite{kumar2018rev2} for our experiments. The directed edge $e(A, B)$ represents the reputation of node B w.r.t. the node A. The positive edges represent trust between users, and the negative edges represent distrust between users. 

\begin{table}[h]
\centering
\begin{tabular}{lllll}\\
\hline
Statistic                    & Value      &  &  &  \\
\hline
Nodes                        & 5881       &  &  &  \\
Edges                        & 35592      &  &  &  \\
Range of Edge weight         & -10 to +10 &  &  &  \\
Percentage of Positive Edges & 89\%       &  &  & \\
\hline
\end{tabular}
\caption {Dataset Statistics}
\end{table}

For ease in experimentation, we reduce the range of edge weights to $-1$ to $+1$ for better interpretation of results.

\section{Background}

\subsection{Link Prediction}
\label{sec-link}
Suppose we are given a graph with three nodes A, B and C. Given that there is a positive relation between A and B, and a negative relation between B and C, and the relation between A and C is not specified. So without link prediction it would be taken as a neutral relation, which might be erroneous. Here, we cannot ignore the fact that according to the Social Balance Theory, the relation between A and C would likely be negative. Thus, link prediction helps in augmenting the signed network and enables it to capture the social interactions robustly.

A simple algorithm suggested in \cite{7837846}, defines two statistics, Fairness and Goodness. The Fairness of a node quantifies how fair a node is, at judging the Goodness of other nodes. The Goodness defines how good or reputed a node actually is. The mutually recursive definitions of Fairness and Goodness of a node are as follows:

\begin{displaymath}
    g(v) = \frac{1}{|in(v)|}\sum_{\substack{u\in{in(v)}}} f(u) * W(u,v)
\end{displaymath}
\begin{displaymath}
    f(u) = 1 -  \frac{1}{|out(u)|}\sum_{\substack{{v\in{out(u)}}}} \frac{|W(u,v) * g(v)|}{R} 
\end{displaymath}
\\ where $in(v)$ is the in-degree of node $v$ and $out(u)$ is the out-degree of node $u$, $W(u, v)$ is the edge weight between node $u$ and $v$, and $R$ is the range normalizing constant. \\
Now, the weight of the edge from a node $u$ to node $v$ is predicted as:
\begin{displaymath}
    E(u,v) = f(u) * g(v)
\end{displaymath}

\subsection{Spectral Clustering}
\label{sec-spec}
Spectral clustering allows us to perform dimensionality reduction on the graph to obtain node embeddings in fewer dimensions. We can subsequently employ clustering techniques on these embeddings to cluster the graph into desired number of clusters.
The motivating idea behind spectral clustering is Graph Drawing. 

\subsubsection{Graph Drawing}
The objective of graph drawing is to generate embeddings for nodes such that a node is close to its neighbours \cite{doi:10.1137/1.9781611972801.49}.

Let $X_i$ $\epsilon$ $\mathbb{R}^2$ be the coordinates of node $i$ in the drawing. For positive weighted edges, we get the following vertex equation:
\begin{displaymath}
X_i = (\sum_{j\sim i}^{} A_{ij})^{-1} \sum_{j\sim i}^{} A_{ij}X_{j}
\end{displaymath}
On simplifying, we obtain, 
\begin{displaymath}
Dx = Ax
\end{displaymath}
\begin{displaymath}
Lx = 0
\end{displaymath}
where $D$ is the diagonal matrix, $A$ is the adjacency matrix and $L = D - A$. We call the L matrix as the unsigned Laplacian Matrix. \\ \\
To extend this equation to Signed Networks, we need to look at some subtleties. First, we want a node to be close to its positive weight neighbours and far from its negative weight neighbours. Second, we need to change the normalising part in the vertex equation. \\
For signed network the equation now becomes:
\begin{displaymath}
X_i = (\sum_{j\sim i}^{} |A_{ij}|)^{-1} \sum_{j\sim i}^{} A_{ij}X_{j}
\end{displaymath}
on simplifying, 
\begin{displaymath}
\bar{D}x = Ax
\end{displaymath}
\begin{displaymath}
\bar{L}x = 0
\end{displaymath}
where $\bar{D}$ is defined as $\bar{D}_{ii} = \sum_{j}^{} |A_{ij}|$ and we obtain the signed Laplacian Matrix, $\bar{L} = \bar{D} - A$. 
\begin{figure}[h]
  \centering
  \includegraphics[width=0.45\textwidth]{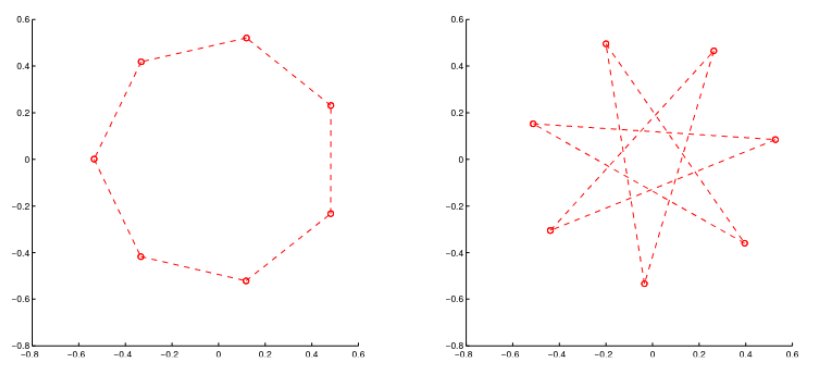}
  \caption{The figure depicts the embeddings for a Graph containing only negative edges by using unsigned Laplacian embeddings and signed Laplacian embeddings respectively from left to right. We see that the nodes are more separated in the right plot. Thus, the signed Laplacian embeddings perform better here \cite{doi:10.1137/1.9781611972801.49}.}
\end{figure}

\subsection{Correlation Clustering}
\label{sec-corr}
Correlation clustering is an algorithm majorly used in a scenario where the relations between the entities of a network are known. The algorithm tries to put similar objects into the same clusters and dissimilar into different clusters.

We use a constant factor approximation algorithm, described in \cite{Bansal2004}, which minimizes the number of Disagreements, i.e., the absolute sum of negative edges inside and the sum of positive edges outside each cluster. This iterative algorithm tries to estimate $\delta$-clean clusters, where $\delta$-clean implies that for each node $v$   $\epsilon$  $C$, the cluster has atleast $(1-\delta)*|C|$ positive-weighted neighbours inside $C$ and atmost $\delta*|C|$ positive-weighted neighbours outside $C$.

One significant advantage of using correlation clustering is that we do not need to fix the optimal number of clusters for our network. The algorithm iteratively determines the best clustering while minimizing the disagreements as defined before. However, the cleanliness parameter ($\delta$) plays a vital role in the clustering performance.

\section{Our Approach}
In this section, we first propose a resiliency metric that helps in capturing the trustworthiness of any individual community in the network. We also devise a few experiments to support the idea of how this metric corresponds to the resilient nature of this network.

\begin{displaymath}
Trustworthiness_{C_i} = \frac{\sum_{edges}^{} w_{(inside +ve)} + \sum_{edges}^{} |w|_{(outside -ve)}}{max(N_{nodes_{C_{i}}}, N_{nodes_{total}} - N_{nodes_{C_{i}}})}
\end{displaymath}

This metric is defined over the basic notion of what is popularly known in these settings as Agreements, which means more no. of positive edges inside and negative edges outside a cluster. 

We also add a normalizing term similar to the normalizing term seen in the formula for the expansion of a set of nodes in a network \cite{arora2009expander}. The intuition behind it is that we want our communities to be large and have higher agreements.

\subsection{Resiliency Computation Pipeline}
\label{exp-1}
For subsequent experiments in this section, $25$\% of the network is considered for ease of experimentation. We first perform link prediction using the Fairness Goodness algorithm as described in \ref{sec-link} on the Bitcoin Trust network. This allows us to capture all the trust relations, that are not explicitly present in the network. We provide the performance evaluation of this algorithm in section \ref{poc1}.
We then use the spectral \ref{sec-spec} and correlation clustering \ref{sec-corr} algorithms to determine all the conflicting communities in the trust network. Further using the trustworthiness metric, the trust level of each community is calculated. In section \ref{poc2}, we also perform a validation experiment to verify the effectuality of signed network clustering.

\begin{table}[]
\centering
\resizebox{\linewidth}{!} {%
\begin{tabular}{ccccc}
\hline
\multicolumn{1}{l}{\textbf{Cluster No.}} & \multicolumn{1}{l}{\textbf{No. of Nodes}} & \multicolumn{1}{l}{\textbf{Inside +ve edges}} & \multicolumn{1}{l}{\textbf{Outside -ve edges}} & \multicolumn{1}{l}{\textbf{Trust Ci}} \\ \hline
1 & 1096 & 2880 & 51 & 0.56469 \\
2 & 355 & 404 & 33 & 0.063114 \\
3 & 299 & 283 & 25 & 0.044304 \\
4 & 49 & 88 & 20 & 0.015648 \\
5 & 1 & 0 & 0 & 0 \\ \hline
\multicolumn{1}{l}{\textbf{Total Clusters:}} & \textbf{5} & \multicolumn{2}{l}{\textbf{Total Trustworthiness of the Network:}} & \textbf{0.687756} \\ \hline
\end{tabular}}
\caption{Spectral clustering results over the original network}
\label{scCT}
\vspace{-4mm}
\end{table}

\begin{table}[]
\centering
\resizebox{\linewidth}{!} {%
\begin{tabular}{ccccc}
\hline
\multicolumn{1}{l}{\textbf{Cluster No.}} & \multicolumn{1}{l}{\textbf{No. of Nodes}} & \multicolumn{1}{l}{\textbf{Inside +ve edges}} & \multicolumn{1}{l}{\textbf{Outside -ve edges}} & \multicolumn{1}{l}{\textbf{Trust Ci}} \\ \hline
1 & 1292 & 5775 & 36 & 0.881811 \\
2 & 96 & 93 & 2 & 0.009331 \\
3 & 15 & 0 & 15 & 0.008403 \\
4 & 97 & 61 & 3 & 0.007692 \\
5 & 99 & 58 & 6 & 0.006232 \\
6 & 33 & 22 & 6 & 0.005037 \\ \hline
\multicolumn{1}{l}{\textbf{Total Clusters:}} & \textbf{27} & \multicolumn{2}{l}{\textbf{Total Trustworthiness of the Network:}} & \multicolumn{1}{l}{\textbf{0.93562}} \\ \hline
\end{tabular}}
\caption{Correlation clustering results over the original network with cleanliness parameter=0.05}
\label{ccCT}
\vspace{-4mm}
\end{table}

The results for spectral and correlation clustering are shown in Tables \ref{scCT} and \ref{ccCT} respectively. The total trustworthiness of the network is defined as the sum of trustworthiness values of all the communities determined using the above approach.
Using the clustering approaches, we were able to detect the best community in the network on the basis of its trustworthiness.

Now, to quantify and analyse the resiliency of this trust network, we perform the following disruption event: We firstly remove the most trustworthy community from the network and then re-evaluate the total trustworthiness of the network by using the similar heuristics as before. All the edge weights are re-calculated using the link prediction algorithm and the trust levels of the community structures are re-determined by applying both the clustering over the disrupted network. The spectral and correlation clustering results over the disrupted network are shown in Table \ref{scDT} and \ref{ccDT} respectively.

We here hypothesize that the change in the total trustworthiness of the network due to the above disruption determines how resilient our trust network is. We give a detailed discussion of these results in section \ref{inferences}.

\begin{table}[]
\centering
\resizebox{\linewidth}{!} {%
\begin{tabular}{ccccc}
\hline
\multicolumn{1}{l}{\textbf{Cluster No.}} & \multicolumn{1}{l}{\textbf{No. of Nodes}} & \multicolumn{1}{l}{\textbf{Inside +ve edges}} & \multicolumn{1}{l}{\textbf{Outside -ve edges}} & \multicolumn{1}{l}{\textbf{Trust Ci}} \\ \hline
1 & 463 & 845 & 17 & 0.37257 \\
2 & 36 & 8 & 5 & 0.007784 \\
3 & 13 & 6 & 4 & 0.005355 \\
4 & 59 & 8 & 1 & 0.003411 \\
5 & 8 & 4 & 1 & 0.002155 \\
6 & 8 & 0 & 1 & 0.001437 \\
7 & 28 & 2 & 0 & 0.000592 \\ \hline
\multicolumn{1}{l}{\textbf{Total Clusters:}} & \textbf{25} & \multicolumn{2}{l}{\textbf{Total Trustworthiness of the Network:}} & \textbf{0.394169} \\ \hline
\end{tabular}}
\caption {Spectral Clustering over the Disrupted Network}
\label{scDT}
\vspace{-4mm}
\end{table}

\begin{table}[]
\centering
\resizebox{\linewidth}{!} {%
\begin{tabular}{ccccc}
\hline
\multicolumn{1}{l}{\textbf{Cluster No.}} & \multicolumn{1}{l}{\textbf{No. of Nodes}} & \multicolumn{1}{l}{\textbf{Inside +ve edges}} & \multicolumn{1}{l}{\textbf{Outside -ve edges}} & \multicolumn{1}{l}{\textbf{Trust Ci}} \\ \hline
1 & 96 & 93 & 2 & 0.038592 \\
2 & 97 & 61 & 1 & 0.031387 \\
3 & 99 & 58 & 5 & 0.023472 \\
4 & 33 & 22 & 4 & 0.015579 \\
5 & 58 & 12 & 2 & 0.009333 \\
6 & 7 & 0 & 4 & 0.005589 \\ \hline
\multicolumn{1}{l}{\textbf{Total Clusters:}} & \textbf{26} & \multicolumn{2}{l}{\textbf{Total Trustworthiness of the Network:}} & \textbf{0.152015} \\ \hline
\end{tabular}}
\caption {Correlation Clustering over the Disrupted Network with cleanliness parameter=0.002}
\label{ccDT}
\vspace{-4mm}
\end{table}

\subsection{Comparison with baseline}
\label{exp-2}
To check for the efficacy of the resiliency scores, we utilise the temporal nature of the Bitcoin OTC network to generate an year-wise distribution of resiliency scores for the network. As a baseline, we also compute resiliency scores using the following method: the disruption event remains the same as in section \ref{exp-1}, but, to compute the resiliency scores, we compare the mean absolute error (MAE) of the link predicted values for the entire network before and after the disruption. 

\begin{figure}[h]
  \centering
  \includegraphics[width=\linewidth * 9/10]{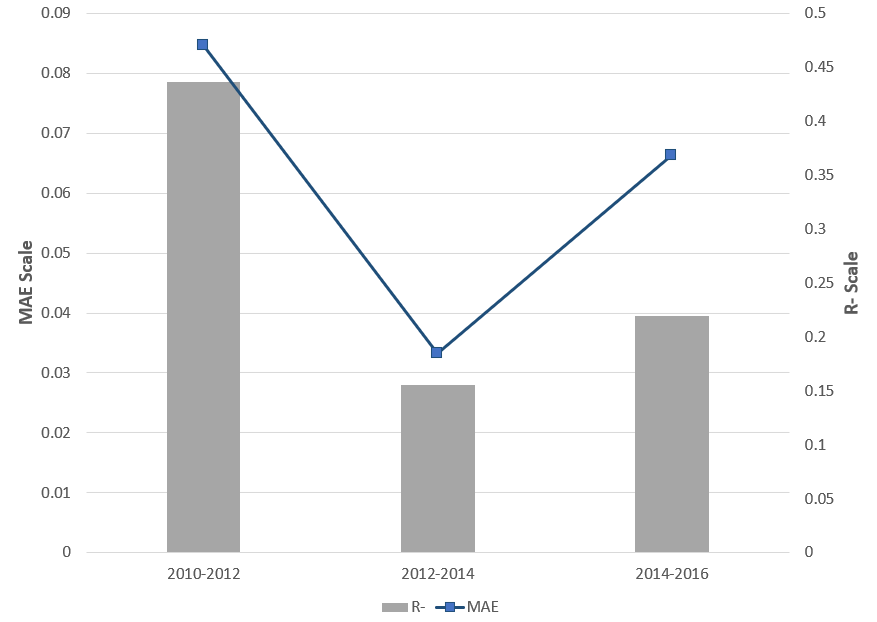}
  \caption{The graph represents the resiliency scores over the years represented by the bar chart. We also show the trend for the MAE scores, represented by the line chart, of the link predicted edges before and after the disruption event. }
    \label{poc}
\end{figure}

We claim that if we see a similar trend in both the techniques for computing resiliency scores over the years, then our hypothesis: the change in total trustworthiness of the network is a good estimate to determine the resiliency of the trust network, is significant. The year-wise distribution of the MAE scores and the change in total trustworthiness of the network ($R^{-}$) due to the disruption event are shown in figure \ref {poc} . We provide a detailed discussion of these results in section \ref{inferences}.

\section{Performance Evaluation}
\subsection{Link Prediction}
\label{poc1}
To evaluate the performance of link prediction algorithm, we use the temporal nature of the Bitcoin OTC Trust Network. This allows us to partition the entire network according to the presence of its entities in a particular time-frame. We evaluate the absolute error between the predicted and the expected edge weights for different split proportions as shown in Table \ref{lpfr}. The average MAE score $0.16$ is quite low, validating that the link prediction is able to effectively capture the unknown edge weights.
\begin{table}[]
\resizebox{\linewidth}{!} {%
\begin{tabular}{ccc}
\hline
\multicolumn{1}{l}{\textbf{Train-Test \%}} & \multicolumn{1}{l}{\textbf{No. of Predicted Edges}} & \multicolumn{1}{l}{\textbf{Mean Absolute Error}} \\ \hline
50\% - 50 \% & 17796 & \textbf{0.1715892} \\
70\% - 30 \% & 10678 & \textbf{0.1869823} \\
90\% - 10 \% & 3560 & \textbf{0.1336905} \\ \hline
\multicolumn{2}{c}{\textbf{Average MAE:}} & \textbf{0.164087} \\ \hline
\end{tabular}}
\caption{Link Prediction results using Fairness and Goodness. The mean absolute error is calculated between the predicted edge weights and the true edge weights detected after the dynamic growth of the network. The network is split on a particular time state based on the train \%.}
\label{lpfr}
\vspace{-4mm}
\end{table}
\subsection{Signed Clustering}
\label{poc2}

To verify that the clustering techniques on signed networks indeed get us a better metric score, we formed a sub-graph from the original graph by only considering the positive edge weights. Thereafter, we applied the spectral clustering on the sub-graph and compared the results with the signed clustering techniques. The results are summarised in table \ref{scpositive}. The metric scores for each type of clustering allow us to inference that the signed clustering techniques benefit us in getting a better metric score. We claim that higher metric score is representative of a better clustering because of the presence of more agreements in a correctly clustered network.
\begin{table}[]
\resizebox{\linewidth}{!} {%
\begin{tabular}{lc}
\hline
\textbf{Spectral Clustering}                     & \multicolumn{1}{l}{\textbf{Total Trustworthiness of the network}} \\ \hline
Only positive edges network                      & \textbf{0.221061}                                                 \\
Both negative \& positive edges (Original Graph) & \textbf{0.687756} \\ \hline                                              
\end{tabular}}
\caption{Total Trustworthiness scores for spectral clustering using only positive edges and using the original graph. }
\label{scpositive}
\end{table}

\section{Inferences}

\label{inferences}

This section includes the discussion on the results from the experiments performed in section \ref{exp-1} and \ref{exp-2}. The tables \ref{scCT} and \ref{ccCT} show the trustworthiness scores of the clusters obtained from spectral and correlation clustering, respectively. We observe that the total trust metric computed by correlation clustering is higher than that of spectral clustering. Through this, we claim that correlation clustering performs better than spectral clustering, as it is able to cluster the graph into better communities, i.e., communities having higher trustworthiness scores.
The tables \ref{scDT} and \ref{ccDT} show the trustworthiness scores of the clusters obtained on the disrupted network described in section \ref{exp-1} using spectral and correlation clustering, respectively. We observe a radical change in the total trustworthiness of the network which reflects that after the disruption event, the strength of the trust network reduces significantly. Thus, we can interpret that the change in the trustworthiness metric is a good estimate to determine the resiliency of this network. Higher the change in the total trustworthiness, lesser the strength and resiliency of the network.

$R^{-} = Trustworthiness_{original} - Trustworthiness_{disrupted}$

To benchmark the performance of the trustworthiness as a metric for resiliency, in \ref{exp-2}, we compare our results with a baseline metric: MAE of the link predicted edge weights of the trust network before and after the disruption event. Figure \ref{poc} shows the trend of  $R^{-}$ and MAE values over the years. We observe a similar trend in both the metrics, which validates that our resiliency metric is concordant with the expected behavior of the network on foreign disruption events.

\section{Conclusion and Future Work}

Disruptions are quite common in real-world networks and could lead to a severe effect on both the life and strength of the network. Thus, resiliency estimation is an important task while analyzing any realistic network. With this work, we present a clustering-based approach to quantify the resiliency of signed social networks by exposing them to a custom disruption event. We investigate the performance of our resiliency metric using various experiments on a real-world signed trust network. Our approach can be extended to assess and quantify the resiliency of various real-world signed networks. The current approach lacks in benchmarking the performance of the resiliency metric on possible realistic disruption events. We would like to devise constrained disruptions based on the level of impact on the network for a more generic assessment. We hope that our work serves as a motivation for others to contribute in the field of resiliency estimation for signed social networks.

\clearpage
\bibliographystyle{ACM-Reference-Format}
\bibliography{sample-base}
\end{document}